\documentclass{article}
\usepackage{spconf,amsmath,amssymb,amsfonts,graphicx,booktabs,multirow,array,tabularx,textcomp,algorithmic,cite,hyperref,cleveref}
\usepackage[table]{xcolor}
\usepackage{orcidlink}
\def\BibTeX{{\rm B\kern-.05em{\sc i\kern-.025em b}\kern-.08em
    T\kern-.1667em\lower.7ex\hbox{E}\kern-.125emX}}

\definecolor{gainhigh}{RGB}{200, 230, 201}
\definecolor{gainlow}{RGB}{245, 245, 245} 

\definecolor{lightgreen}{RGB}{225, 245, 225}
\definecolor{lightgray}{RGB}{240, 240, 240}
\definecolor{highlightgreen}{HTML}{C6EFCE}
\definecolor{goldhighlight}{HTML}{FFE699}

\definecolor{colIntra}{HTML}{D6EADF} 
\definecolor{colSea}{HTML}{D0E6EB}   
\definecolor{colInter}{HTML}{FCE3D2} 

\newcolumntype{W}{>{\centering\arraybackslash}p{9mm}}    
\newcolumntype{O}{>{\centering\arraybackslash}p{18mm}}   

\newcommand{\arw}[1]{\makebox[0pt][l]{$\boldsymbol{#1}$}\hspace{0.62em}}
\newcommand{\pct}[2]{%
  \makebox[11mm][l]{\color{gray}\scriptsize\textit{(\arw{#1}#2\%)}}}

\newcommand{\orthd}[2]{\makebox[8mm][c]{\textbf{#1}}\,\pct{\downarrow}{#2}}
\newcommand{\orthu}[2]{\makebox[8mm][c]{#1}\,\pct{\uparrow}{#2}}

\title{Language Orthogonalization for\\Zero-Shot Cross-Lingual Audio Deepfake Detection}
\name{Minu Kim\orcidlink{0009-0005-4520-5071}$^{1,2}$, Ji Sub Um$^{2}$, Hoirin Kim$^{2}$}
\address{$^{1}$University of Southern California, USA, $^{2}$KAIST, South Korea\\
\small\texttt{minukim@usc.edu, \{twiz0311,hoirkim\}@kaist.ac.kr}}
\begin{document}
%
\maketitle

\begin{abstract}
Audio deepfake detectors need to transfer to languages absent from training, as multilingual speech synthesis outpaces labeled anti-spoofing resources. While detectors increasingly rely on self-supervised speech models (S3Ms), these backbones encode language-dependent structure that confounds spoof cues. We address this confound through language orthogonalization, a target-free ridge map that removes S3M variation projected onto continuous language-identification (LID) embeddings. Across six languages, six S3M backbones, and all Leave-$N$-Out settings, it consistently reduces EER across unseen languages. Cross-lingual EER correlates with LID-space distance, where orthogonalization yields larger gains for more distant transfers.
\end{abstract}

\section{Introduction}
\label{sec:intro}

Consider deploying an audio deepfake detector in a language entirely absent from training: neither bonafide nor spoofed speech is available for adaptation or calibration. This scenario is increasingly plausible: speech synthesis now spans over a thousand languages~\cite{pratap2024scaling}, while standard anti-spoofing benchmarks remain largely English-centric~\cite{todisco2019asvspoof,liu2022asvspoof} and recent multilingual resources cover only a fraction of this breadth~\cite{muller2024mlaad}. A scalable detector should therefore transfer both its bonafide reference and spoof decision boundary across languages.

Self-supervised speech models (S3Ms) appear well suited to this transfer setting, having shown strong performance in audio deepfake detection~\cite{tak2022automatic}. However, their representations also encode language identity~\cite{liu2022efficient}, and spoof detectors often degrade across linguistic boundaries~\cite{marek2024audio,borodin2026spoof}. A detector may easily confuse language-conditioned acoustics with transferable spoof cues. Because both bonafide and spoofed speech retain this linguistic structure, cross-lingual degradation is especially severe when transferring to distant target languages.

Meanwhile, prior work shows that S3M phonetic variations occupy approximately linear subspaces~\cite{choi2026self,choi2026b}, motivating subspace residualization to reduce cross-lingual confounds~\cite{kim2026language}. We hypothesize that removing linear language directions eliminates confounds across large linguistic distances without target data. We thus formulate \textit{language orthogonalization} for audio deepfake detection under a strictly unseen setting, removing language variation without observing target-language speech during training.

\begin{figure*}[t]
\centering
\includegraphics[width=0.95\textwidth]{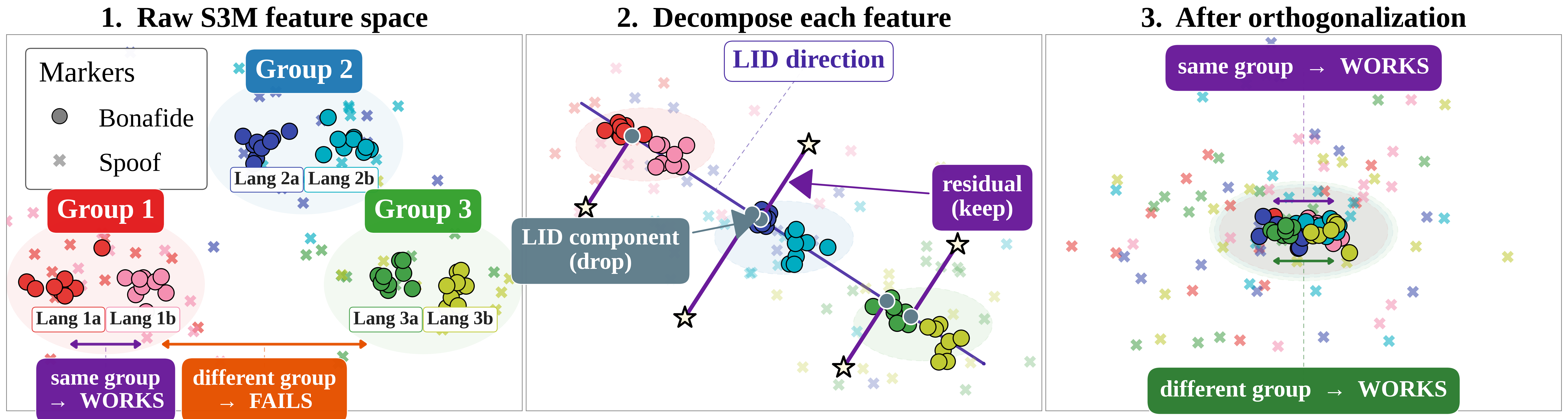}
\caption{\textbf{Language orthogonalization for cross-lingual anti-spoofing.} (1) S3M representations place similar languages nearby, easing transfer between close languages. (2) Orthogonalization removes language-predictable variation and retains the residual. (3) The residual space aligns bonafide distributions and improves transfer across linguistic groups.}
\label{fig}
\end{figure*}

Our approach fits a closed-form ridge map from continuous language identification (LID) embeddings to S3M representations using bonafide speech from source languages, then subtracts the mapped language component from each representation. A frozen multilingual LID encoder~\cite{valk2021voxlingua107} extracts continuous language representations without language-specific tuning, enabling target-data-free transformation. Consequently, this orthogonalization isolates and preserves spoof artifacts while removing language confounds.

By leveraging continuous LID space, which provides a meaningful geometry reflecting phonological, geographic, and genealogical language relationships~\cite{kim2026scaling,kim2025improving}, we first show that cross-lingual equal error rate (EER) increases with LID distance across six South and Southeast Asian languages and six S3M backbones. Language orthogonalization consistently improves Leave-$N$-Out evaluation across all settings, with the most pronounced gains on distant transfers where language confounds are most severe.

Our contributions are: (1) showing that LID distance and linguistic composition govern cross-lingual transfer difficulty; (2) formulating target-free language orthogonalization for unseen-language audio deepfake detection; and (3) demonstrating consistent EER reductions across six S3Ms and all Leave-$N$-Out settings, with largest gains on distant transfers and constrained source setups.

\begin{figure}[t]
\centering
\includegraphics[width=0.95\columnwidth]{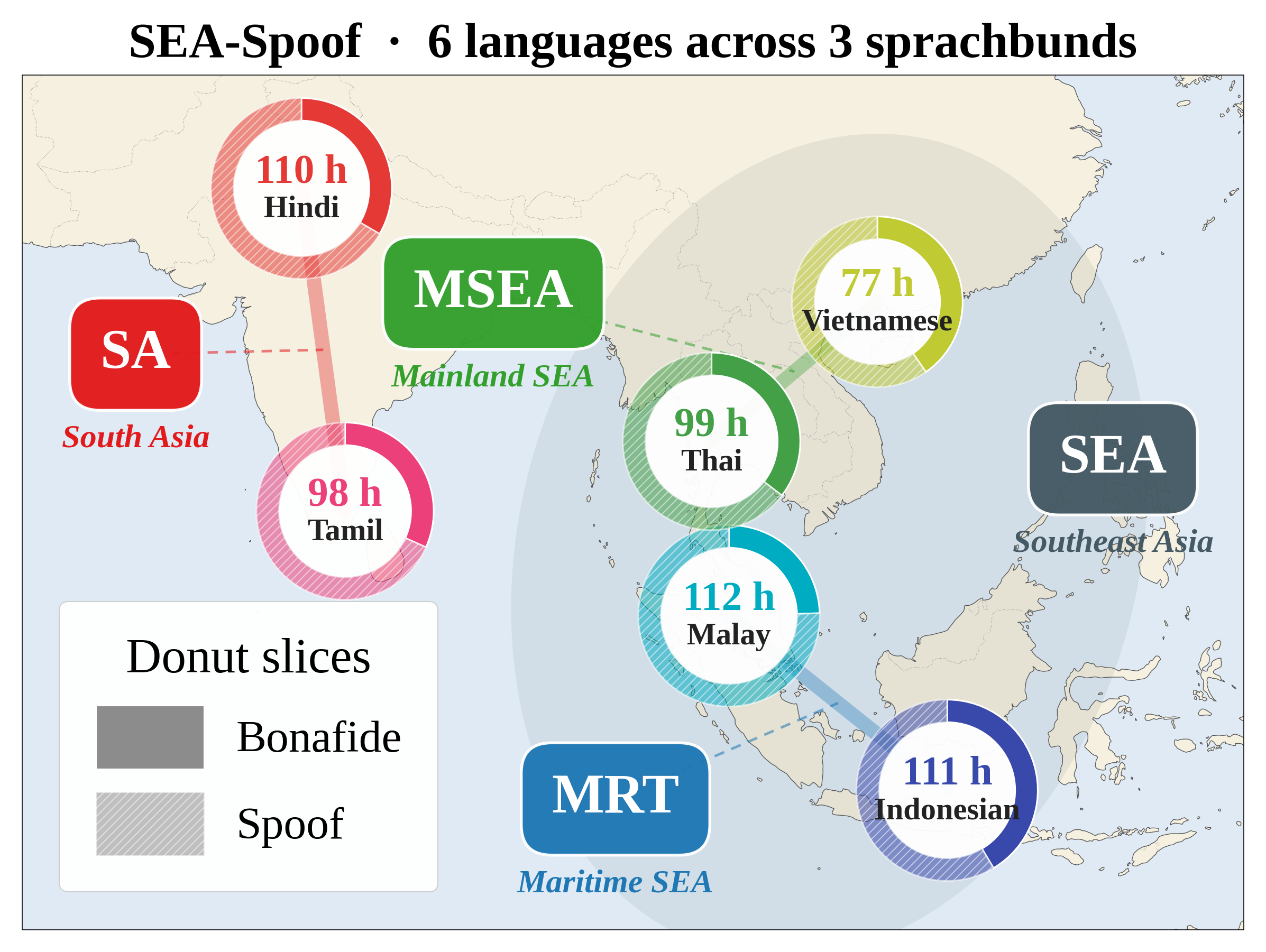}
\caption{\textbf{Evaluation languages.} Six languages span South Asia, Mainland Southeast Asia, and Maritime Southeast Asia, enabling transfer within and across linguistic areas (i.e., sprachbunds).}
\label{fig:dataset}
\end{figure}

\section{Methodology}
\label{sec:method}

\subsection{Feature Representations}
\label{sec}

We use six frozen S3Ms as utterance-level feature extractors: MMS-300M~\cite{pratap2024scaling}, XLS-R-300M~\cite{babu2021xls}, wav2vec2-Large-LV60~\cite{baevski2020wav2vec}, HuBERT-Large~\cite{hsu2021hubert}, WavLM-Large~\cite{chen2022wavlm}, and mHuBERT-147~\cite{boito2024mhubert}. For each utterance, we compute mean+std pooling~\cite{klempir2026statistical} across all hidden layers, average the layer-wise representations depthwise, and $\ell_2$-normalize the result to obtain a 2048-d embedding $\mathbf{x}_i \in \mathbb{R}^{D}$. This fixed pipeline avoids backbone-specific layer selection~\cite{pasad2021layer,pasad2023comparative}.

As an external language reference, we extract a continuous LID embedding $\mathbf{g}_i\in\mathbb{R}^{256}$ using an ECAPA-TDNN pretrained on VoxLingua107~\cite{valk2021voxlingua107}. Unlike discrete language labels, these embeddings place utterances in a shared language space, enabling both language-distance analysis and transformation of target languages unseen during training.

\subsection{Language Orthogonalization}
\label{subsec:orth}

We propose a source-only formulation of language orthogonalization for fully unseen-language anti-spoofing. The mapping is learned exclusively from source-language bonafide speech, requiring neither class from target languages and preventing spoof artifacts from entering its estimation.

For the bonafide training set $\mathcal{D}_{\mathrm{bf}}$, we stack its $N_{\mathrm{bf}} = |\mathcal{D}_{\mathrm{bf}}|$ utterances into S3M features $\mathbf{X}_{\mathrm{bf}} \in \mathbb{R}^{N_{\mathrm{bf}}\times 2048}$ and LID embeddings $\mathbf{G}_{\mathrm{bf}} \in \mathbb{R}^{N_{\mathrm{bf}}\times 256}$ to fit a ridge map:

\begin{equation}
\mathbf{W}^{*} = \arg\min_{\mathbf{W}}
\left\|
\mathbf{X}_{\mathrm{bf}}-\mathbf{G}_{\mathrm{bf}}\mathbf{W}
\right\|_F^2
+\lambda\|\mathbf{W}\|_F^2 .
\label{eq:ridge_objective}
\end{equation}

The closed-form solution is

\begin{equation}
\mathbf{W}^{*} =
\left(
\mathbf{G}_{\mathrm{bf}}^{\top}\mathbf{G}_{\mathrm{bf}}
+\lambda\mathbf{I}
\right)^{-1}
\mathbf{G}_{\mathrm{bf}}^{\top}\mathbf{X}_{\mathrm{bf}} .
\label{eq:closed_form}
\end{equation}

For an utterance with S3M representation $\mathbf{X}$ and LID embedding
$\mathbf{g}$, we subtract the language-predictable component:

\begin{equation}
\mathbf{X}_{\mathrm{orth}} =
\mathbf{X}-
\mathbf{g}\mathbf{W}^{*}.
\label{eq:orth}
\end{equation}

The transformation is applied to both bonafide and spoofed utterances. Its continuous LID conditioning allows application to target languages unseen during training.\footnote{We report main results at $\lambda = 0.1$ as a representative setting, with performance remaining highly stable across $\lambda \in (0, 1)$.}

\begin{figure*}[t]
\centering
\includegraphics[width=0.85\textwidth]{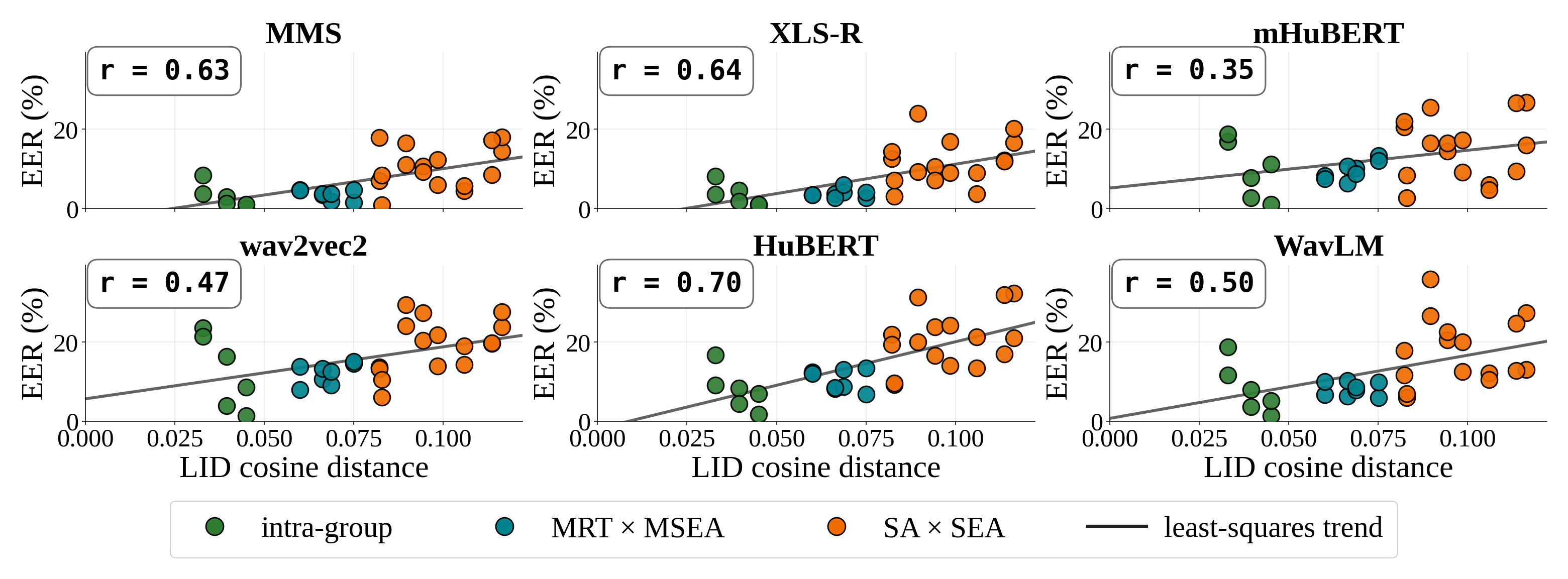}
\caption{\textbf{Language distance predicts cross-lingual difficulty.}
LID distances reflect linguistic-area structure, with intra-group pairs closer than MSEA--MRT and SEA--SA pairs. Across six S3Ms, greater distance is consistently associated with higher EER; $r$ denotes the Pearson correlation between distance and EER.}
\label{fig:language_distance}
\end{figure*}

\begin{table*}[t]
\centering
\setlength{\tabcolsep}{4pt}
\caption{\textbf{Cross-lingual EER (\%) under Leave-$N$-Out evaluation.}
The remaining $6-N$ languages are used for training. Parentheses show
relative EER reduction from Raw. Orthogonalization improves every backbone
and $N$.}
\label{tab:leave_n_sweep}
\resizebox{\linewidth}{!}{%
\begin{tabular}{l | W O | W O | W O | W O | W O}
\toprule
& \multicolumn{2}{c|}{\textbf{$N=1$}} 
& \multicolumn{2}{c|}{\textbf{$N=2$}} 
& \multicolumn{2}{c|}{\textbf{$N=3$}} 
& \multicolumn{2}{c|}{\textbf{$N=4$}} 
& \multicolumn{2}{c}{\textbf{$N=5$}} \\
\textbf{Backbone}
& Raw & Orth
& Raw & Orth
& Raw & Orth
& Raw & Orth
& Raw & Orth \\
\midrule
\multicolumn{11}{l}{\textit{Multilingual Pre-trained Backbones}} \\
\midrule
MMS-300M & \cellcolor{lightgray}1.48 & \cellcolor{lightgreen}\orthd{1.16}{22} & \cellcolor{lightgray}2.44 & \cellcolor{lightgreen}\orthd{1.82}{25} & \cellcolor{lightgray}3.54 & \cellcolor{lightgreen}\orthd{2.68}{24} & \cellcolor{lightgray}5.65 & \cellcolor{lightgreen}\orthd{3.91}{31} & \cellcolor{lightgray}10.06 & \cellcolor{lightgreen}\orthd{7.80}{22} \\
XLS-R-300M & \cellcolor{lightgray}1.62 & \cellcolor{lightgreen}\orthd{1.58}{2} & \cellcolor{lightgray}2.36 & \cellcolor{lightgreen}\orthd{2.14}{9} & \cellcolor{lightgray}3.59 & \cellcolor{lightgreen}\orthd{2.89}{20} & \cellcolor{lightgray}6.36 & \cellcolor{lightgreen}\orthd{4.24}{33} & \cellcolor{lightgray}11.56 & \cellcolor{lightgreen}\orthd{7.77}{33} \\
mHuBERT-147 & \cellcolor{lightgray}5.16 & \cellcolor{lightgreen}\orthd{4.39}{15} & \cellcolor{lightgray}6.46 & \cellcolor{lightgreen}\orthd{5.49}{15} & \cellcolor{lightgray}7.95 & \cellcolor{lightgreen}\orthd{7.07}{11} & \cellcolor{lightgray}10.56 & \cellcolor{lightgreen}\orthd{9.37}{11} & \cellcolor{lightgray}16.10 & \cellcolor{lightgreen}\orthd{14.76}{8} \\
\midrule
\multicolumn{11}{l}{\textit{Monolingual Pre-trained Backbones}} \\
\midrule
wav2vec2-large & \cellcolor{lightgray}6.45 & \cellcolor{lightgreen}\orthd{5.75}{11} & \cellcolor{lightgray}8.50 & \cellcolor{lightgreen}\orthd{7.23}{15} & \cellcolor{lightgray}10.34 & \cellcolor{lightgreen}\orthd{9.05}{12} & \cellcolor{lightgray}12.86 & \cellcolor{lightgreen}\orthd{11.44}{11} & \cellcolor{lightgray}17.20 & \cellcolor{lightgreen}\orthd{15.63}{9} \\
HuBERT-large & \cellcolor{lightgray}5.47 & \cellcolor{lightgreen}\orthd{5.29}{3} & \cellcolor{lightgray}6.84 & \cellcolor{lightgreen}\orthd{6.65}{3} & \cellcolor{lightgray}9.10 & \cellcolor{lightgreen}\orthd{8.70}{4} & \cellcolor{lightgray}12.66 & \cellcolor{lightgreen}\orthd{11.30}{11} & \cellcolor{lightgray}17.54 & \cellcolor{lightgreen}\orthd{15.54}{11} \\
WavLM-large & \cellcolor{lightgray}4.32 & \cellcolor{lightgreen}\orthd{4.18}{3} & \cellcolor{lightgray}5.93 & \cellcolor{lightgreen}\orthd{5.23}{12} & \cellcolor{lightgray}7.83 & \cellcolor{lightgreen}\orthd{6.72}{14} & \cellcolor{lightgray}10.86 & \cellcolor{lightgreen}\orthd{8.86}{18} & \cellcolor{lightgray}15.56 & \cellcolor{lightgreen}\orthd{13.14}{16} \\
\midrule
Avg. & \cellcolor{lightgray}4.08 & \cellcolor{lightgreen}\orthd{3.73}{9} & \cellcolor{lightgray}5.42 & \cellcolor{lightgreen}\orthd{4.76}{12} & \cellcolor{lightgray}7.06 & \cellcolor{lightgreen}\orthd{6.19}{12} & \cellcolor{lightgray}9.83 & \cellcolor{lightgreen}\orthd{8.19}{17} & \cellcolor{lightgray}14.67 & \cellcolor{lightgreen}\orthd{12.44}{15} \\
\bottomrule
\end{tabular}%
}
\end{table*}

\section{Experimental Setup}
\label{sec:experimental}

\subsection{Dataset}
\label{subsec:dataset}

We evaluate on SEA-Spoof~\cite{wu2025sea}, which contains
bonafide and spoofed speech from six languages under diverse speakers and spoofing conditions. We group the languages into South Asia (SA; Hindi and Tamil), Mainland Southeast Asia (MSEA; Thai and Vietnamese), and Maritime Southeast Asia (MRT; Indonesian and Malay) (Fig.~\ref{fig:dataset}); MSEA and MRT together form the Southeast Asian (SEA) subset. Prior S3M studies likewise report representational proximity and effective
transfer among languages in these regions~\cite{kim2026far,kim2026scaling},
motivating comparisons within and across linguistic areas.

\subsection{Cross-Lingual Evaluation}
\label{subsec:cross_lingual_evaluation}

Under Leave-$N$-Out evaluation, we hold out every combination of
$N\in\{1,2,3,4,5\}$ languages and train on the remaining $6-N$. The held-out
languages are entirely unseen, with no language-specific adaptation. For each
split, we compare raw and orthogonalized S3M representations using EER, averaged across all language combinations at each $N$.
To evaluate the representations, we fit a logistic regression classifier with class-balanced $L_2$ regularization using L-BFGS~\cite{pedregosa2011scikit}.

\section{Results}
\label{sec:results}

\subsection{Language Distance and Cross-Lingual Difficulty}
\label{subsec:language_distance}

For each source--target language pair, we compare the cosine distance of their mean LID embeddings against cross-lingual evaluation EER. Figure~\ref{fig:language_distance} shows that LID distance reflects linguistic-area structure and correlates with EER across all six S3Ms. Transfer difficulty thus aligns with linguistic distance from the training language, motivating orthogonalization under severe linguistic distance.

\subsection{Overall Leave-$N$-Out Performance}
\label{subsec:cross_lingual_results}

Table~\ref{tab:leave_n_sweep} reports performance as unseen languages increase from $N=1$ to $5$. Orthogonalization consistently reduces EER across all backbones and $N$, yielding 9--17\% average relative gains across all setups.

\subsection{Effect of Linguistic-Area Composition}
\label{subsec:area_composition}

We examine how linguistic-area structure influences cross-lingual difficulty and orthogonalization gains under data scarcity. Under single-language training (Table~\ref{tab:orth_effect_set2}), transfer difficulty scales with linguistic mismatch: Raw baseline performance is strong on closely related pairs (\texttt{intra-group}), leaving limited room for improvement. Conversely, Raw performance drops on distant cross-area pairs ($\mathrm{SA} \times \mathrm{SEA}$) due to severe linguistic confounds. Crucially, orthogonalization yields its largest gains on these distant pairs (e.g., up to 22\% relative EER reduction on XLS-R), confirming that language removal is most effective across large linguistic distances.

Under two-language training (Table~\ref{tab:orth_effect_set1}), generalization depends on source diversity. Diverse training pairs ($\mathrm{SA} \times \mathrm{SEA}$) yield robust baselines with low Raw EERs, leading to minor orthogonalization gains. In contrast, homogeneous pairs (\texttt{intra-group}) restrict linguistic coverage, degrading baseline accuracy. Orthogonalization markedly improves these constrained setups (e.g., up to 41\% and 45\% EER reduction on XLS-R for \texttt{intra-group} and \texttt{MRT $\times$ MSEA}), showing that removing LID-mapped components can effectively compensate for limited source diversity.

\begin{table}[t]
\centering
\setlength{\tabcolsep}{4pt}
\caption{\textbf{Cross-lingual EER (\%) under single-language training (Leave-5-Out).} Raw results are shown in \colorbox{lightgray}{gray};
Orth results are grouped by source--target relation:
\colorbox{colIntra}{\texttt{intra-group}},
\colorbox{colSea}{\texttt{MRT $\times$ MSEA}}, and
\colorbox{colInter}{\texttt{SA $\times$ SEA}}.
Parentheses indicate relative EER change from Raw.}
\label{tab:orth_effect_set2}
\resizebox{\columnwidth}{!}{
\begin{tabular}{l | W O | W O | W O}
\toprule & \multicolumn{6}{c}{\textit{Source--target relation at evaluation}} \\
\cmidrule(lr){2-7} & \multicolumn{2}{c|}{\textbf{\texttt{intra-group}}} & \multicolumn{2}{c|}{\textbf{\texttt{MRT $\times$ MSEA}}} & \multicolumn{2}{c}{\textbf{\texttt{SA $\times$ SEA}}} \\ 
\cmidrule(lr){2-3} \cmidrule(lr){4-5} \cmidrule(lr){6-7} 
\textbf{Backbone} & Raw & Orth & Raw & Orth & Raw & Orth \\ 
\midrule 
\multicolumn{7}{l}{\textit{Multilingual Pre-trained Backbones}} \\ 
\midrule
MMS & \cellcolor{lightgray}2.90 & \cellcolor{colIntra}\orthd{2.89}{0} & \cellcolor{lightgray}3.54 & \cellcolor{colSea}\orthu{3.70}{5} & \cellcolor{lightgray}10.52 & \cellcolor{colInter}\orthd{8.62}{18} \\
XLS-R & \cellcolor{lightgray}3.33 & \cellcolor{colIntra}\orthu{4.05}{22} & \cellcolor{lightgray}3.75 & \cellcolor{colSea}\orthd{3.43}{9} & \cellcolor{lightgray}11.72 & \cellcolor{colInter}\orthd{9.18}{22} \\
mHuBERT & \cellcolor{lightgray}9.72 & \cellcolor{colIntra}\orthd{8.29}{15} & \cellcolor{lightgray}9.63 & \cellcolor{colSea}\orthu{10.40}{8} & \cellcolor{lightgray}15.16 & \cellcolor{colInter}\orthd{15.14}{0} \\
\midrule
\multicolumn{7}{l}{\textit{Monolingual Pre-trained Backbones}} \\
\midrule
wav2vec2 & \cellcolor{lightgray}12.53 & \cellcolor{colIntra}\orthd{11.51}{8} & \cellcolor{lightgray}12.13 & \cellcolor{colSea}\orthd{11.78}{3} & \cellcolor{lightgray}19.04 & \cellcolor{colInter}\orthd{17.35}{9} \\
HuBERT & \cellcolor{lightgray}7.85 & \cellcolor{colIntra}\orthd{7.65}{3} & \cellcolor{lightgray}10.41 & \cellcolor{colSea}\orthd{9.67}{7} & \cellcolor{lightgray}20.42 & \cellcolor{colInter}\orthd{18.26}{11} \\
WavLM & \cellcolor{lightgray}8.09 & \cellcolor{colIntra}\orthd{6.54}{19} & \cellcolor{lightgray}8.17 & \cellcolor{colSea}\orthd{7.51}{8} & \cellcolor{lightgray}17.56 & \cellcolor{colInter}\orthd{15.76}{10} \\
\bottomrule
\end{tabular}%
}
\end{table}

\begin{table}[t]
\centering
\setlength{\tabcolsep}{4pt}
\caption{\textbf{Cross-lingual EER (\%) under two-language training (Leave-4-Out).} Raw results are shown in \colorbox{lightgray}{gray};
Orth results are grouped by training-language composition:
\colorbox{colIntra}{\texttt{intra-group}},
\colorbox{colSea}{\texttt{MRT $\times$ MSEA}}, and
\colorbox{colInter}{\texttt{SA $\times$ SEA}}.
Parentheses indicate relative EER reduction from Raw.}
\label{tab:orth_effect_set1}
\resizebox{\columnwidth}{!}{
\begin{tabular}{l | W O | W O | W O} 
\toprule & \multicolumn{6}{c}{\textit{Composition of the two training languages}} \\
\cmidrule(lr){2-7} & \multicolumn{2}{c|}{\textbf{\texttt{intra-group} train}} & \multicolumn{2}{c|}{\textbf{\texttt{MRT $\times$ MSEA} train}} & \multicolumn{2}{c}{\textbf{\texttt{SA $\times$ SEA} train}} \\ 
\cmidrule(lr){2-3} \cmidrule(lr){4-5} \cmidrule(lr){6-7} 
\textbf{Backbone} & Raw & Orth & Raw & Orth & Raw & Orth \\ 
\midrule
\multicolumn{7}{l}{\textit{Multilingual Pre-trained Backbones}} \\
\midrule
MMS & \cellcolor{lightgray}8.66 & \cellcolor{colIntra}\orthd{5.91}{32} & \cellcolor{lightgray}6.86 & \cellcolor{colSea}\orthd{3.85}{44} & \cellcolor{lightgray}3.91 & \cellcolor{colInter}\orthd{3.19}{18} \\
XLS-R & \cellcolor{lightgray}10.82 & \cellcolor{colIntra}\orthd{6.44}{41} & \cellcolor{lightgray}9.67 & \cellcolor{colSea}\orthd{5.30}{45} & \cellcolor{lightgray}3.03 & \cellcolor{colInter}\orthd{2.90}{5} \\
mHuBERT & \cellcolor{lightgray}14.56 & \cellcolor{colIntra}\orthd{13.15}{10} & \cellcolor{lightgray}13.05 & \cellcolor{colSea}\orthd{10.10}{23} & \cellcolor{lightgray}7.82 & \cellcolor{colInter}\orthd{7.58}{3} \\
\midrule
\multicolumn{7}{l}{\textit{Monolingual Pre-trained Backbones}} \\
\midrule
wav2vec2 & \cellcolor{lightgray}16.24 & \cellcolor{colIntra}\orthd{14.30}{12} & \cellcolor{lightgray}15.64 & \cellcolor{colSea}\orthd{13.35}{15} & \cellcolor{lightgray}10.20 & \cellcolor{colInter}\orthd{9.42}{8} \\
HuBERT & \cellcolor{lightgray}17.67 & \cellcolor{colIntra}\orthd{16.22}{8} & \cellcolor{lightgray}16.55 & \cellcolor{colSea}\orthd{13.41}{19} & \cellcolor{lightgray}8.83 & \cellcolor{colInter}\orthd{8.41}{5} \\
WavLM & \cellcolor{lightgray}14.40 & \cellcolor{colIntra}\orthd{12.41}{14} & \cellcolor{lightgray}13.99 & \cellcolor{colSea}\orthd{10.53}{25} & \cellcolor{lightgray}7.96 & \cellcolor{colInter}\orthd{6.70}{16} \\
\bottomrule
\end{tabular}%
}
\end{table}

\begin{figure}[t]
    \centering
    \includegraphics[width=\columnwidth]{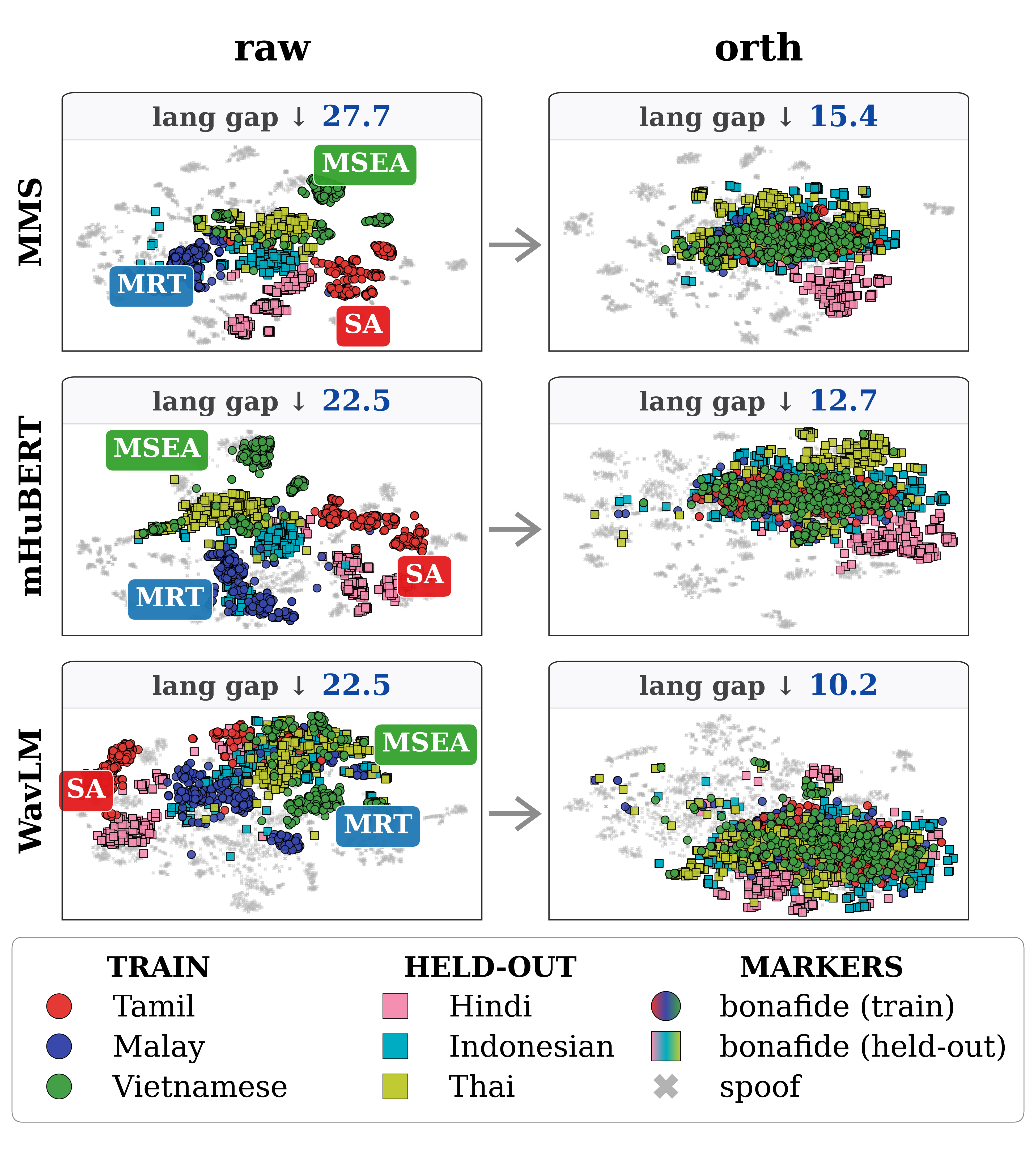}
    \caption{\textbf{Representation geometry before and after orthogonalization.} Raw representations cluster hierarchically by language and linguistic group, whereas orthogonalization aligns bonafide speech and improves bonafide--spoof separation. Reported values denote Euclidean distances between bonafide language centroids in the S3M feature space.}
    \label{fig:tsne}
\end{figure}

\subsection{Representation Geometry after Orthogonalization}
\label{subsec:representation_geometry}

Figure~\ref{fig:tsne} visualizes the representation geometry in a representative Leave-3-Out setting. Raw S3M representations cluster hierarchically by language and linguistic group, with substantial domain shifts across languages. Orthogonalization suppresses this language-dependent structure, effectively aligning bonafide representations across both source and unseen target languages into a shared space.

Quantitatively, orthogonalization substantially reduces the Euclidean distance between bonafide language centroids in the S3M feature space across all backbones: MMS ($27.7 \to 15.4$, $-44\%$), mHuBERT ($22.5 \to 12.7$, $-44\%$), WavLM ($22.5 \to 10.2$, $-55\%$), XLS-R ($20.3 \to 9.4$, $-54\%$), wav2vec 2.0 ($21.2 \to 6.4$, $-70\%$), and HuBERT ($23.4 \to 9.7$, $-59\%$). This spatial reorganization confirms that removing LID-predictable components closes cross-lingual domain gaps while retaining the discriminative representation needed for detection.

\section{Conclusion}
\label{sec:conclusion}

We introduce a language orthogonalization strategy that fits an LID-to-S3M mapping on bonafide speech to improve cross-lingual audio deepfake detection. Across multiple S3M backbones and unseen-language setups, this approach consistently lowers EER, yielding larger gains on linguistically distant transfers. Furthermore, we show that continuous distance in LID space reflects linguistic-area structure, providing a clear metric to characterize transfer difficulty.

\vfill\pagebreak

\newpage

\footnotesize
\section{Acknowledgments}
This work was supported by Institute of Information \& communications Technology Planning \& Evaluation (IITP) grant funded by the Korea government(MSIT) (No.RS-2025-02215393).

\bibliographystyle{IEEEbib}
\begingroup
\bibliography{strings,refs}

@article{pratap2024scaling,
  title={Scaling speech technology to 1,000+ languages},
  author={Pratap, Vineel and Tjandra, Andros and Shi, Bowen and Tomasello, Paden and Babu, Arun and Kundu, Sayani and Elkahky, Ali and Ni, Zhaoheng and Vyas, Apoorv and Fazel-Zarandi, Maryam and others},
  journal={Journal of Machine Learning Research},
  volume={25},
  number={97},
  pages={1--52},
  year={2024}
}

@article{todisco2019asvspoof,
  title={ASVspoof 2019: Future horizons in spoofed and fake audio detection},
  author={Todisco, Massimiliano and Wang, Xin and Vestman, Ville and Sahidullah, Md and Delgado, H{\'e}ctor and Nautsch, Andreas and Yamagishi, Junichi and Evans, Nicholas and Kinnunen, Tomi and Lee, Kong Aik},
  journal={arXiv preprint arXiv:1904.05441},
  year={2019}
}

@article{liu2022asvspoof,
  title={Asvspoof 2021: Towards spoofed and deepfake speech detection in the wild},
  author={Liu, Xuechen and Wang, Xin and Sahidullah, Md and Patino, Jose and Delgado, H{\'e}ctor and Kinnunen, Tomi and Todisco, Massimiliano and Yamagishi, Junichi and Evans, Nicholas and Nautsch, Andreas and others},
  journal={arXiv preprint arXiv:2210.02437},
  year={2022}
}

@inproceedings{muller2024mlaad,
  title={Mlaad: The multi-language audio anti-spoofing dataset},
  author={M{\"u}ller, Nicolas M and Kawa, Piotr and Choong, Wei Herng and Casanova, Edresson and G{\"o}lge, Eren and M{\"u}ller, Thorsten and Syga, Piotr and Sperl, Philip and B{\"o}ttinger, Konstantin},
  booktitle={2024 International Joint Conference on Neural Networks (IJCNN)},
  pages={1--7},
  year={2024},
  organization={IEEE}
}

@article{marek2024audio,
  title={Are audio DeepFake detection models polyglots?},
  author={Marek, Bart{\l}omiej and Kawa, Piotr and Syga, Piotr},
  journal={arXiv preprint arXiv:2412.17924},
  year={2024}
}

@article{borodin2026spoof,
  title={When spoof detectors travel: Evaluation across 66 languages in the low-resource language spoofing corpus},
  author={Borodin, Kirill and Kudryavtsev, Vasiliy and Maslov, Maxim and Gorodnichev, Mikhail and Mkrtchian, Grach},
  journal={arXiv preprint arXiv:2603.02364},
  year={2026}
}

@inproceedings{tak2022automatic,
  title     = {{Automatic Speaker Verification Spoofing and Deepfake Detection Using Wav2vec 2.0 and Data Augmentation}},
  author    = {Hemlata Tak and Massimiliano Todisco and Xin Wang and Jee-weon Jung and Junichi Yamagishi and Nicholas Evans},
  year      = {2022},
  booktitle = {{The Speaker and Language Recognition Workshop (Odyssey 2022)}},
  pages     = {112--119},
  doi       = {10.21437/Odyssey.2022-16},
}

@article{liu2022efficient,
  title={Efficient self-supervised learning representations for spoken language identification},
  author={Liu, Hexin and Perera, Leibny Paola Garcia and Khong, Andy WH and Chng, Eng Siong and Styles, Suzy J and Khudanpur, Sanjeev},
  journal={IEEE Journal of Selected Topics in Signal Processing},
  volume={16},
  number={6},
  pages={1296--1307},
  year={2022},
  publisher={IEEE}
}

@inproceedings{choi2026b,
  title={[b]=[d]-[t]+[p]: Self-supervised Speech Models Discover Phonological Vector Arithmetic},
  author={Choi, Kwanghee and Yeo, Eunjung and Cho, Cheol Jun and Harwath, David and Mortensen, David R},
  booktitle={Findings of the Association for Computational Linguistics: ACL 2026},
  pages={11048--11069},
  year={2026}
}

@article{choi2026self,
  title={Self-supervised speech models encode phonetic context via position-dependent orthogonal subspaces},
  author={Choi, Kwanghee and Yeo, Eunjung and Cho, Cheol Jun and Mortensen, David R and Harwath, David},
  journal={arXiv preprint arXiv:2603.12642},
  year={2026}
}

@inproceedings{valk2021voxlingua107,
  title={{VoxLingua107}: A dataset for spoken language recognition},
  author={Valk, J{\"o}rgen and Alum{\"a}e, Tanel},
  booktitle={Proc. IEEE SLT},
  pages={652--658},
  year={2021}
}

@inproceedings{kim2025improving,
  title={Improving cross-lingual phonetic representation of low-resource languages through language similarity analysis},
  author={Kim, Minu and Jang, Kangwook and Kim, Hoirin},
  booktitle={ICASSP 2025-2025 IEEE International Conference on Acoustics, Speech and Signal Processing (ICASSP)},
  pages={1--5},
  year={2025},
  organization={IEEE}
}

@inproceedings{kim2026far,
  title={How Far Do SSL Speech Models Listen for Tone? Temporal Focus of Tone Representation under Low-Resource Transfer},
  author={Kim, Minu and Um, Ji Sub and Kim, Hoirin},
  booktitle={ICASSP 2026-2026 IEEE International Conference on Acoustics, Speech and Signal Processing (ICASSP)},
  pages={18297--18301},
  year={2026},
  organization={IEEE}
}

@article{kim2026scaling,
  title={Scaling Self-Supervised Speech Models Uncovers Deep Linguistic Relationships: Evidence from the Pacific Cluster},
  author={Kim, Minu and Kim, Hoirin and Mortensen, David R},
  journal={arXiv preprint arXiv:2603.07238},
  year={2026}
}

@article{babu2021xls,
  title={XLS-R: Self-supervised cross-lingual speech representation learning at scale},
  author={Babu, Arun and Wang, Changhan and Tjandra, Andros and Lakhotia, Kushal and Xu, Qiantong and Goyal, Naman and Singh, Kritika and Von Platen, Patrick and Saraf, Yatharth and Pino, Juan and others},
  journal={arXiv preprint arXiv:2111.09296},
  year={2021}
}

@article{baevski2020wav2vec,
  title={wav2vec 2.0: A framework for self-supervised learning of speech representations},
  author={Baevski, Alexei and Zhou, Yuhao and Mohamed, Abdelrahman and Auli, Michael},
  journal={Advances in neural information processing systems},
  volume={33},
  pages={12449--12460},
  year={2020}
}

@article{hsu2021hubert,
  title={Hubert: Self-supervised speech representation learning by masked prediction of hidden units},
  author={Hsu, Wei-Ning and Bolte, Benjamin and Tsai, Yao-Hung Hubert and Lakhotia, Kushal and Salakhutdinov, Ruslan and Mohamed, Abdelrahman},
  journal={IEEE/ACM transactions on audio, speech, and language processing},
  volume={29},
  pages={3451--3460},
  year={2021},
  publisher={IEEE}
}

@article{chen2022wavlm,
  title={Wavlm: Large-scale self-supervised pre-training for full stack speech processing},
  author={Chen, Sanyuan and Wang, Chengyi and Chen, Zhengyang and Wu, Yu and Liu, Shujie and Chen, Zhuo and Li, Jinyu and Kanda, Naoyuki and Yoshioka, Takuya and Xiao, Xiong and others},
  journal={IEEE Journal of Selected Topics in Signal Processing},
  volume={16},
  number={6},
  pages={1505--1518},
  year={2022},
  publisher={IEEE}
}

@article{boito2024mhubert,
  title={mhubert-147: A compact multilingual hubert model},
  author={Boito, Marcely Zanon and Iyer, Vivek and Lagos, Nikolaos and Besacier, Laurent and Calapodescu, Ioan},
  journal={arXiv preprint arXiv:2406.06371},
  year={2024}
}

@inproceedings{pasad2021layer,
  title={Layer-wise analysis of a self-supervised speech representation model},
  author={Pasad, Ankita and Chou, Ju-Chieh and Livescu, Karen},
  booktitle={2021 IEEE Automatic Speech Recognition and Understanding Workshop (ASRU)},
  pages={914--921},
  year={2021},
  organization={IEEE}
}

@inproceedings{pasad2023comparative,
  title={Comparative layer-wise analysis of self-supervised speech models},
  author={Pasad, Ankita and Shi, Bowen and Livescu, Karen},
  booktitle={ICASSP 2023-2023 IEEE International Conference on Acoustics, Speech and Signal Processing (ICASSP)},
  pages={1--5},
  year={2023},
  organization={IEEE}
}

@article{wu2025sea,
  title={SEA-Spoof: Bridging The Gap in Multilingual Audio Deepfake Detection for South-East Asian},
  author={Wu, Jinyang and Hou, Nana and Pan, Zihan and Zhang, Qiquan and Bhupendra, Sailor Hardik and Mondal, Soumik},
  journal={arXiv preprint arXiv:2509.19865},
  year={2025}
}

@article{kim2026language,
  title={Language Orthogonalization of Self-Supervised Speech Representations for Cross-lingual Parkinson's Detection},
  author={Kim, Minu and Yeo, Eunjung and Choi, Kwanghee and Kim, June-Woo},
  journal={arXiv preprint arXiv:2609.09499},
  year={2026}
}

@article{klempir2026statistical,
  title={Statistical, multi-scale and attention-based layer pooling of Wav2Vec-2 speech embeddings for Parkinson's disease detection},
  author={Klempir, Ondrej and Mullerova, Juliana Grand and Krupicka, Radim},
  journal={Computers in Biology and Medicine},
  volume={200},
  pages={111368},
  year={2026},
  publisher={Elsevier}
}

@article{pedregosa2011scikit,
  title={Scikit-learn: Machine learning in Python},
  author={Pedregosa, Fabian and Varoquaux, Ga{\"e}l and Gramfort, Alexandre and Michel, Vincent and Thirion, Bertrand and Grisel, Olivier and Blondel, Mathieu and Prettenhofer, Peter and Weiss, Ron and Dubourg, Vincent and others},
  journal={the Journal of machine Learning research},
  volume={12},
  pages={2825--2830},
  year={2011},
  publisher={JMLR. org}
}

\end{document}